\documentclass[10pt,letterpaper]{article}
\usepackage[top=0.85in,left=2.75in,footskip=0.75in,marginparwidth=2in]{geometry}

\usepackage[utf8]{inputenc}

\usepackage{cite}

\usepackage{nameref,hyperref}

\usepackage[right]{lineno}

\usepackage{microtype}
\DisableLigatures[f]{encoding = *, family = * }

\raggedright
\usepackage{changepage}

\usepackage[aboveskip=1pt,labelfont=bf,labelsep=period,singlelinecheck=off]{caption}

\makeatletter
\renewcommand{\@biblabel}[1]{\quad#1.}
\makeatother

\usepackage{lastpage,fancyhdr,graphicx}
\usepackage{epstopdf}
\fancyheadoffset[L]{2.25in}
\fancyfootoffset[L]{2.25in}

\usepackage{color}

\definecolor{Gray}{gray}{.25}

\usepackage{graphicx}

\usepackage{sidecap}

\usepackage{wrapfig}
\usepackage[pscoord]{eso-pic}
\usepackage[fulladjust]{marginnote}
\reversemarginpar

\title{\bfseries{Freestanding dielectric nanohole array metasurface for mid-infrared wavelength applications}}
\begin{document}
	
	\maketitle
	
\author  {Jun Rong Ong,$^{1,+}$ }
\author {Hong Son Chu,$^{1}$ }
\author {Valerian Hongjie Chen,$^{1,2}$ }
\author {Alexander Yutong Zhu, $^{3}$}
\author {Patrice Genevet$^{4,*}$\\}
\bigskip 
{$^1$}  Institute of High Performance Computing, 1 Fusionopolis Way, 16-16 Connexis, 138632, Singapore \\
{$^2$}  Rudolf Peierls Centre for Theoretical Physics, University of Oxford, Oxford, UK\\
{$^3$}  John A. Paulson School of Engineering and Applied Sciences, Harvard University, Cambridge, Massachusetts 02138, United States\\
{$^4$}  Universite Cote d'Azur, CNRS, CRHEA, rue Bernard Gregory, Sophia Antipolis 06560 Valbonne, France\\

\bfseries{$^*$} Corresponding author: patrice.genevet@crhea.cnrs.fr\\

\begin{abstract}
We designed and simulated freestanding dielectric optical metasurfaces based on arrays of etched nanoholes in a silicon membrane. We showed $2\pi$ phase control and high forward transmission at mid-infrared wavelengths around $4.2$ $\mu$m by tuning the dimensions of the holes. We also identified the mechanisms responsible for high forward scattering efficiency and showed that these conditions are connected with the well-known Kerker conditions already proposed for isolated scatterers. A beam deflector was designed and optimized through sequential particle swarm and gradient descent optimization to maximize transmission efficiency and reduce unwanted grating orders. Such freestanding silicon nanohole array metasurfaces are promising for the realization of silicon based mid-infrared optical elements. 
\end{abstract}

\thispagestyle{fancy}

Metasurfaces are two-dimensional planar metamaterials formed through suitable arrangement of etched subwavelength structures. They have the inherent advantage of being thin, lightweight compared to traditional optics and yet straightforward to fabricate compared to three-dimensional metamaterials \cite{yu2011light,yu2014flat,genevet2017recent,lin2014dielectric,kildishev2013planar,decker2015high}. Inheriting concepts proposed in pioneering works on high contrast gratings and microwave reflect and transmit-arrays, metasurface technology is now suitable for applications at optical wavelengths \cite{kock1948metallic,stork1991artificial,farn1992binary,lalanne1998blazed,lalanne1999design,berry1963reflectarray,pozar1993analysis,huang2007reflectarray}. Engineered optical metasurfaces have the ability to impart an abrupt phase change to the incident wavefront over a subwavelength thickness by utilizing structural resonances, enabling many potential applications in beam deflection, lensing and wavefront control. 

\begin{figure}[ht]
	\centering
	\includegraphics[width=0.8\columnwidth,trim= 5cm 4cm 5cm 4cm, clip]{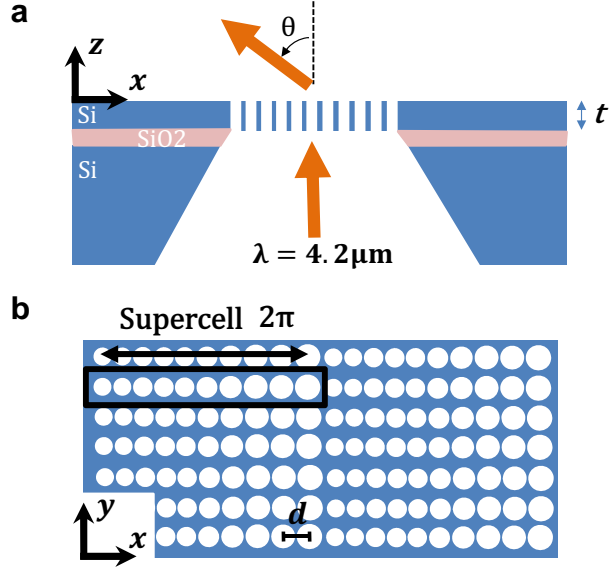}
	\caption{(a) Schematic of freestanding silicon nanohole array metasurface acting as a beam deflector with deflect angle $\theta$. (b) Top-down view of silicon nanohole array beam deflector metasurface. The metasurface is constructed using repeating supercells. Within each supercell, the holes are separated by a center-to-center distance $d$. From first hole to last hole there is a linear phase gradient and the end-to-end the phase shift difference is $2\pi$. }
	\label{fig:1}
\end{figure}

Mid-IR photonics at wavelengths of 2 to $20$ $\mu$m has wide ranging applications in spectroscopy, chemical and biomolecular sensing, and detection \cite{lee2007widely,soref2010mid}. Metasurface optics at mid-IR wavelengths could potentially be simultaneously highly transparent, easy to manufacture and of low-cost. However, fabricating a metasurface on a supporting substrate material may impose an additional undesirable material constraint on the intended application, e.g. high absorption of silicon dioxide substrate in the mid-IR wavelengths. A freestanding and transmitting silicon based metasurface is thus a highly attractive optical element at mid-IR wavelengths. In this article, we report the design of a freestanding and transmitting beam deflector made using an arrangement of subwavelength holes in a silicon membrane (see Fig. \ref{fig:1}). Similarly configured silicon based freestanding nanohole arrays have previously been reported in the literature\cite{ho2014development,park2014subwavelength}. However, these nanohole arrays were either designed as reflectors or were unintentionally strongly reflecting. In our deflector, after a traditional parametric study, we further performed optimization of the hole dimensions and positions using particle swarm and gradient descent techniques to maximize forward transmission and suppress unwanted grating orders. 

\begin{figure}[ht]
	\centering
	\includegraphics[width=1\columnwidth,trim= 2cm 6cm 2cm 5cm, clip]{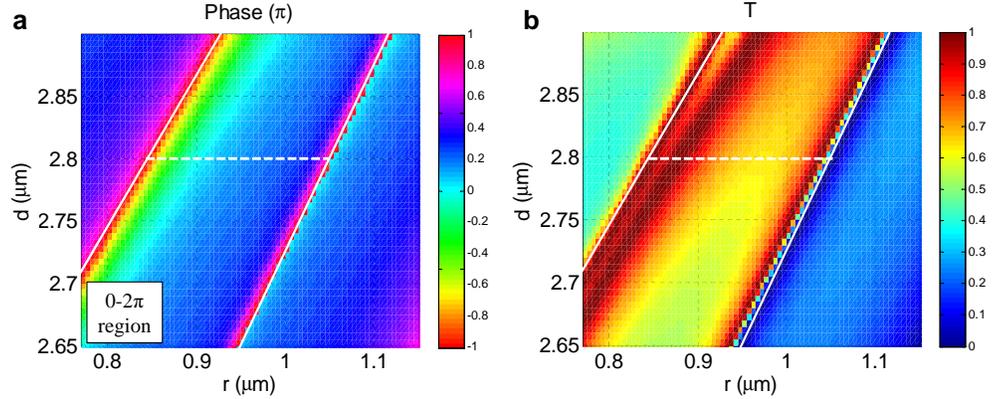}
	\caption{3D FDTD simulation results for (a) phase and (b) transmission $T$ of nanohole arrays at $r = 770$ to 1150 nm and $d = 2650$ to 2900 nm. The solid white line demarcates the region with full $2\pi$ range of phase shift. The dotted white line indicates a hole period $d$ of 2800 nm which was used in subsequent simulations.}
	\label{fig:2}
\end{figure}

We first consider the phase change and transmission through a periodic square lattice of circular holes in a freestanding silicon membrane. We performed 3D finite difference time domain (FDTD) simulations \cite{lumerical} while changing the hole radius $r$ and hole period $d$ and fixing the slab thickness at $t = 1.118$ $\mu$m. The silicon nanohole array is surrounded by air ($n=1$). The excitation source is chosen to be plane waves polarized along the x-axis (parallel to the periodicity) and normally incident on one side of the silicon. Figure \ref{fig:2} shows the simulation results for transmission and phase at $r = 770$ to 1150 nm and $d = 2650$ to 2900 nm. For many potential applications of metasurfaces, phase control over a full 2$\pi$ is needed. From Fig. \ref{fig:2}(a), we can identify the region of interest where $2\pi$ phase control is achievable, as demarcated by the solid white line borders. Correspondingly, Fig. \ref{fig:2}(b) shows the transmission levels for the dimensions within this region of interest. Based on these simulation results, we choose a hole period $d$ of 2800 nm (dotted line)for our subsequent calculations so as to have $T>0.6$ for all hole radii $r$ within the region. 

\begin{figure}[ht]
	\centering
	\includegraphics[width=1\columnwidth,trim= 1cm 4cm 1cm 5cm, clip]{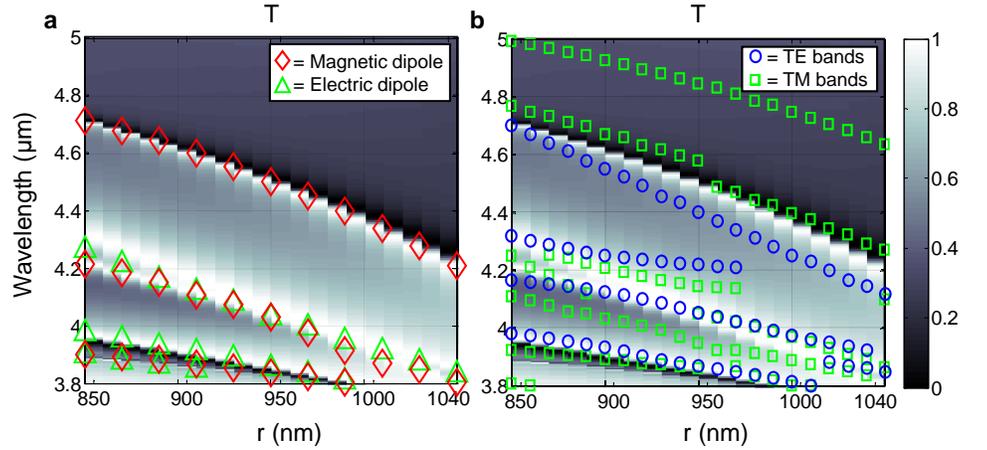}
	\caption{3D FDTD simulation results for transmission $T$ vs. wavelength for $r = 850$ to 1040 nm. In (a),the peaks of the electric and magnetic dipolar contributions to the scattering cross section are overlayed showing good correlation with high forward transmission when they are coincident. In (b), the bandstructure of the nanohole array at the $\Gamma$-point in k-space is overlayed.}
	\label{fig:3}
\end{figure}

It has been shown both theoretically and experimentally that in-phase interference of electric and magnetic dipole modes, i.e. Kerker conditions, in high index dielectric nanoparticles can produce high forward light scattering \cite{kerker1983electromagnetic,geffrin2012magnetic,fu2013directional,alaee2015generalized}. To show that the high forward transmission $T$ of the nanohole array is due to the fulfillment of Kerker conditions, we performed a multipole decomposition of the fields being excited within the structure by normally incident plane waves \cite{yu2015high,miroshnichenko2015nonradiating,campione2015tailoring}. The electromagnetic fields were extracted from our FDTD simulations and decomposed into the electric and magnetic dipole contributions to the total scattering cross section. We consider nanoholes in a periodic square lattice where the lattice interactions are taken into account by the periodic boundary conditions of our simulations. In Fig. \ref{fig:3} we plot the normalized forward transmission spectrum for holes of different radii $r$. In Fig. \ref{fig:3}(a), we overlay the peaks in the electric and magnetic dipole contributions. We found good correlations between coincident dipole contributions and high forward transmission, particularly along the diagonal running from $\lambda=4.2$ $\mu$m and $r = 850$ nm to $\lambda=3.8$ $\mu$m and $r = 1040$ nm. This allows us to infer that there is fulfillment of Kerker conditions within these regions. 

As the structure is a periodic array of nanoholes etched in a membrane, it can also be analyzed as a photonic crystal slab. In Fig. \ref{fig:3}(b), we overlay the TE and TM bandstructure of the square lattice hole array at the $\Gamma$-point ($k_\parallel=0$). We again see correlation between coincident bands and high forward transmission. The bandstructure was calculated using FDTD simulations. Note that these bands are leaky modes above the light line and hence can radiate out of the slab and contribute to the forward transmission. We surmise that the coincidence of these $\Gamma$-point modes is also an indication of the fulfillment of Kerker conditions. 

We designed a mid-IR beam deflector at $4.2$ $\mu$m as a demonstration of the functionality of such freestanding nanohole array metasurfaces. To form a beam deflector, we chose the spatial variation of the hole radii such as to impart a phase shift onto the incoming wave which varies linearly with distance in the x-direction, as in Fig. \ref{fig:1}. The deflection angle $\theta$ of the wavefront is given by the linear phase gradient of the deflector metasurface, 
\begin{equation}
\theta = \sin^{-1}\frac{2\pi}{kNd}
\end{equation}
where a supercell of $N$ number of holes is formed when the total phase shift difference is $2\pi$, $d$ is the hole spacing and $k$ is the free space wave number. A supercell can then be repeated to provide the linear phase gradient over the entire structured metasurface. As a start, we allow the hole spacing $d$ to remain constant over the entire design.  For the current deflector design, we have chosen a repeating supercell of 10 nanoholes. To form the supercell, we have 5 different chosen hole radii $r(\textrm{nm}) = [849, 863, 890, 1021, 1044]$, i.e. adjacent pairs of holes will be of the same radii. The holes are arranged in order of increasing hole radii going from left to right. The deflection angle obtained is $\theta = \sin^{-1}\frac{0.4\pi}{k(2d)} = 8.63^{\circ}$. The hole radiii w chosen purely by the phase gradient requirement and not all radii will have high transmission according to Fig. \ref{fig:2} and \ref{fig:3}. We therefore further optimized the positions and radii of the holes as described in later section. 

\begin{figure}[ht]
	\centering
	\includegraphics[width=1\columnwidth,trim= 4cm 2cm 4cm 4cm, clip]{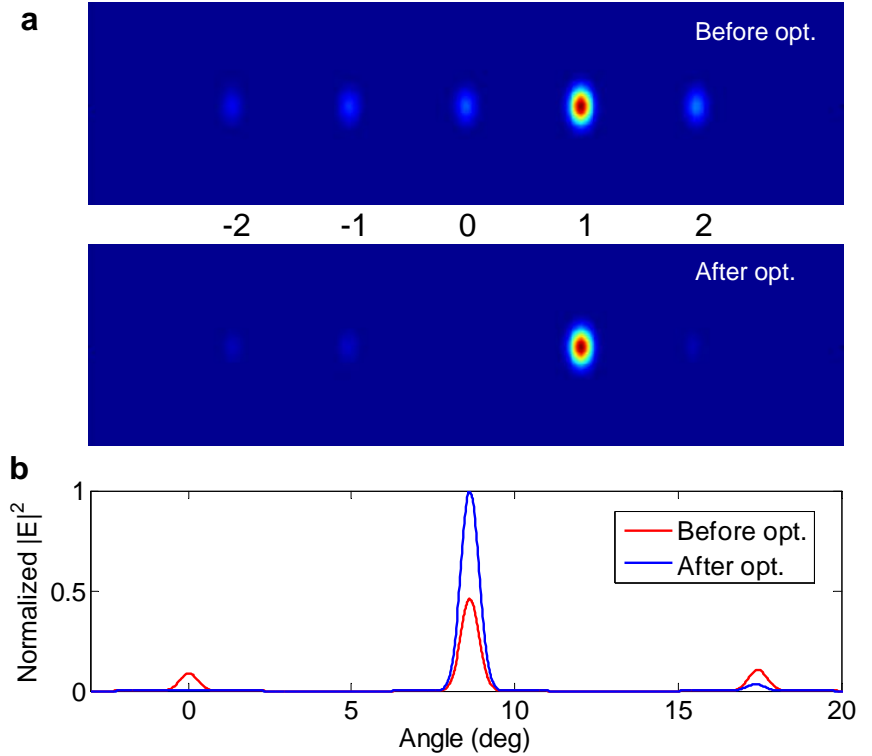}
	\caption{(a) Normalized far field radiation pattern of nanohole array beam deflector, before and after optimization of hole sizes and positions. The numbers indicate the grating orders $m$. (b) Normalized far field radiation pattern, showing stronger scattering into the main lobe and suppression of unwanted grating lobes after optimization.}
	\label{fig:5}
\end{figure}

In order to verify the beam deflecting property of the designed freestanding metasurface, we calculated the far field scattering pattern from the 3D FDTD results (see Fig. \ref{fig:5}(a)). The 3D FDTD simulation was performed with a plane wave source linearly polarized along the long axis of the supercell (x-axis). The strong peak at $\theta = 8.63^{\circ}$ shows the metasurface can efficiently deflect the incident light with good transmission and directivity. The far field pattern also reveals the presence of grating order peaks at angles given by the diffraction grating equation
\begin{equation}
\theta = \sin^{-1}\frac{m \lambda}{Nd}
\end{equation}
where $N$ is the number of holes in the supercell, the order $m = 0$, and $\frac{m\lambda}{Nd}<1$. These peaks reduce the overall efficiency of the beam deflector since not all scattered energy is directed towards a single desired peak. Moreover, we find there is some residual reflection which reduces overall transmission. In the following section, we describe how the beam deflector efficiency was optimized.

To improve the maximum beam deflector efficiency at $4.2$ $\mu$m, we propose to tune the supercell hole radii and also their positions along the left-right axis in order to maximize a figure-of-merit (FOM) \cite{lalanne1998blazed,lalanne1999design}. Similar methods are also commonly used to optimize linear antenna arrays and in designs of metasurfaces \cite{khodier2005linear,zhao2017fast,egorov2017genetically}.  In total, we have 10 radius parameters $r_1$ to $r_{10}$ and 10 position parameters $\Delta d_1$ to $\Delta d_{10}$. The position parameters $\Delta d_i$ denote the hole positions relative to perfect periodicity, i.e. the $i$-th shifted hole position will be $(i-1)\times d + \Delta d_i$. We adopted a sequential particle swarm optimization (PSO) and gradient descent (GD) optimization coupled with FDTD simulations to tune the 20 parameters. The PSO is used to initially search a large number of candidate solutions \cite{robinson2004particle}. Once a pre-determined number of iterations is completed, a GD is performed using the solution with the best FOM as a starting position. As such we are able to arrive at the local optimum around the best position as determined by the PSO. We choose a FOM such as to maximize the forward transmission into the main lobe at the desired deflection angle. Explicitly, the FOM is calculated as
\begin{equation}
\textrm{FOM} = T\times \frac{\int_{\theta_1}^{\theta_2}|E(\theta,\phi_0)|^2 d\theta}{\int_{-\pi/2}^{\pi/2}|E(\theta,\phi_0)|^2 d\theta}
\end{equation}
which is the product of the normalized forward transmission and the fraction that is directed at the main lobe of the deflector. $E(\theta,\phi)$ is the far field on the hemisphere, $\theta_1$ and $\theta_2$ are chosen to be at the midpoints to the adjacent grating orders and $\phi_0 = 0^{\circ}$ is aligned parallel to the x-axis.

The minimum FDTD mesh size was set to 20 nm within the volume of the silicon slab. Symmetric boundary conditions were used along the long edge of the supercell, periodic boundary conditions were used in the short edge of the supercell and in the z-direction PML boundary conditions were used. We set the number of generations in the PSO algorithm to 50 and the number of particles to 20. The PSO was then repeated 3 times, each time re-centering and shrinking the parameter range around the best parameters found from the previous PSO run. Subsequently, we used the best parameters obtained from the final PSO as the starting point for the GD optimization which was done for 30 iterations, each iteration consisting of 21 simulations. We monitored the transmission and reflection of the metasurface and also calculated the near to far field projection. For the optimization, the FOM was calculated at a single wavelength of $4.2$ $\mu$m. 

\begin{figure}[ht]
	\centering
	\includegraphics[width=1\columnwidth,trim= 2cm 2cm 3.5cm 2cm, clip]{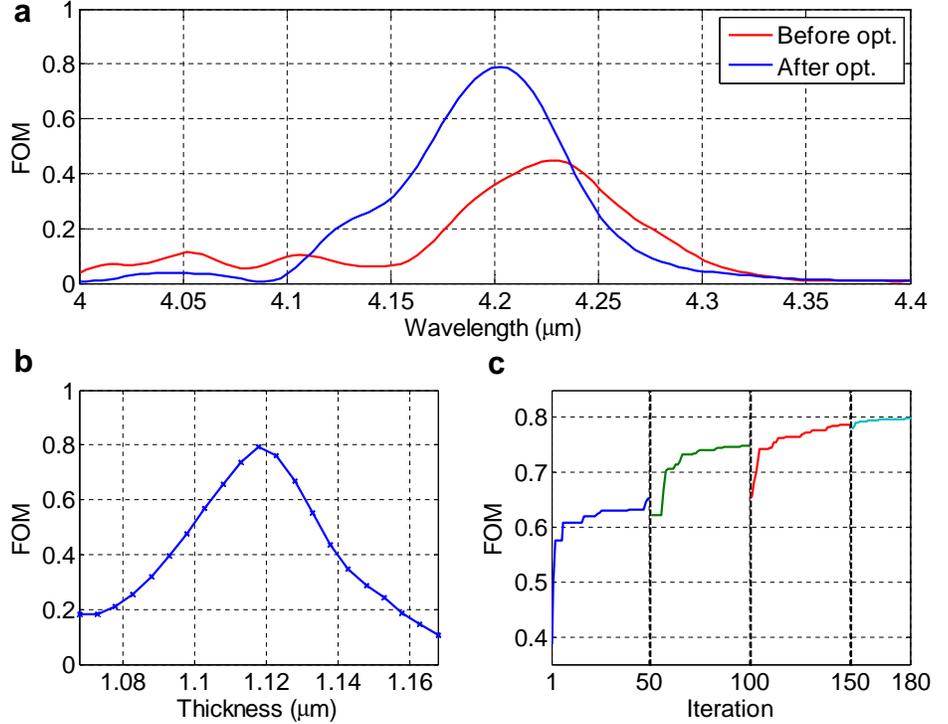}
	\caption{(a) FOM vs. wavelength, before and after optimization. (b) FOM vs. silicon membrane thickness. (c) FOM vs. optimization iteration number. }
	\label{fig:6}
\end{figure}

Figure \ref{fig:5}(a) compares the normalized far field pattern before and after the sequential PSO and GD optimization. We can see that the unwanted grating orders are reduced after optimization. In Fig. \ref{fig:5}(b) we have normalized the far fields to the maximum of the peak obtained after optimization so as to make relative comparisons. The main lobe is shown to have doubled and the adjacent grating orders are reduced. After  optimization, the FOM was increased from 0.365 to 0.795, as shown in Fig. \ref{fig:6}(a). Also, the normalized forward transmission $T$ increased from 0.64 to 0.92. This compares very well to an unpatterned silicon membrane in which $T$ is about 0.6. The final parameters used for the solution with the best FOM are listed in Table \ref{tab:1}. 

\begin{table}[ht]
	\centering
	\begin{tabular}{|c|c||c|c|}
		\hline
		Radii & (nm) & Shift & (nm)\\
		\hline
		$r_1$ & 949 & $\Delta d_1$ & 1\\
		\hline
		$r_2$ & 807 & $\Delta d_2$ & 26\\
		\hline
		$r_3$ & 854 & $\Delta d_3$ & -16\\
		\hline
		$r_4$ & 894 & $\Delta d_4$ & -7\\
		\hline
		$r_5$ & 851 & $\Delta d_5$ & -3\\
		\hline
		$r_6$ & 847 & $\Delta d_6$ & -76\\
		\hline
		$r_7$ & 916 & $\Delta d_7$ & -77\\
		\hline
		$r_8$ & 1033 & $\Delta d_8$ & 74\\
		\hline
		$r_9$ & 987 & $\Delta d_9$ & 26\\
		\hline
		$r_{10}$ & 1120 & $\Delta d_{10}$ & 29\\
		\hline
	\end{tabular}
	\caption{\label{tab:1} Optimized parameters.}
\end{table}

We studied the spectral bandwidth of operation of our nanohole array metasurface deflector. The FWHM of the FOM after optimization is 80 nm (see Fig. \ref{fig:6}(a)), which is sufficiently broad for applications in integrated laser collimators and lenses. Prior to optimization, the actual peak FOM is 0.447 located at $4.23$ $\mu$m. After optimization, the peak was raised and shifted to $4.2$ $\mu$m. We also studied the sensitivity of the optimized FOM to silicon membrane thickness as shown in Fig. \ref{fig:6}(b). At approximately $\pm 25$ nm, the FOM falls to half the maximum. For future work, we can also take into consideration the design sensitivity to thickness through robust optimization methods \cite{men2014robust}. We have plotted the change in FOM with optimization iteration number in Fig. \ref{fig:6}(c) for reference. 

A practical concern of such photonic device optimization algorithms that utilise full 3D FDTD simulations to calculate the FOM is the computational cost of the large number of iterations required. For the optimization procedure described in the above section, each simulation took on average 14 minutes to complete. The total time needed to complete the sequential PSO and GD optimization is close to 850 hours. This large computational time becomes prohibitive when designing multiple complex optical elements. A possible method to reduce the total time required would be to perform an initial coarse mesh optimization followed by a refined mesh optimization. By halving the mesh size on all three dimensions in the FDTD simulation domain, the total simulation time can be reduced by a factor of 8. We verified that our coarse mesh simulation was indeed completed in 1.75 minutes. Subsequently, we completed the PSO optimization with a coarse mesh arriving at a best FOM of 0.791. After mesh refinement, the best FOM was reduced to 0.774 which verifies that a coarse mesh optimization could be used to reduce computation time. 

In this work, we have designed a freestanding silicon nanohole array beam deflector for the mid-IR wavelength of $4.2$ $\mu$m. Our silicon metasurface can be fabricated with conventional techniques and has high transmission due to the lack of substrate, fulfillment of Kerker conditions and subsequent further optimization. Similar principles can be used to design various other optical metasurfaces, e.g. flat lenses. Rectangular, elliptical, or other asymmetrical hole shapes can be used to impart birefringence, as required in waveplates or polarisation beam splitters. Hole shapes with handedness can give chiral optical properties, such as in optical rotation. In the future, we expect our nanohole array design to be able to form useful metasurface optical elements at wavelengths spanning from visible to infrared. 

\section*{Funding Information}

  P. G. gratefully acknowledges financial support from the European Research Council (ERC) under the European Union’s Horizon 2020 research and innovation programme (grant agreement FLATLIGHT No. 639109). J.R.O., H.S.C. and V.H.C. acknowledge the funding support from Agency for Science, Technology and Research - Science and Engineering Research Council for Pharos grant award No. 152-73-00025.


\bibliographystyle{abbrv}

\end{document}